\documentclass{article}
\usepackage{dcase2023,graphicx,url,times,booktabs, tabularx}
\usepackage{url, times, booktabs, tabularx, amssymb, bm, bbm}
\usepackage{algorithm, algorithmic}

\usepackage{array}
\usepackage{enumitem}
\usepackage{soul}
\usepackage{cite}
\usepackage{siunitx}
\usepackage{adjustbox}
\usepackage{hhline}
\usepackage{makecell}
\usepackage{caption}
\usepackage{float}
\usepackage{steinmetz}
\usepackage{tablefootnote}
\usepackage{scrextend}
\usepackage{multirow}

\title{META-SELD: Meta-Learning for Fast Adaptation to the new environment in Sound Event Localization and Detection}

\name{Jinbo Hu$^{1,2}$, 
    Yin Cao$^{3}$, 
    Ming Wu$^{1}$, 
    Feiran Yang$^{1}$, 
    Ziying Yu$^{1}$,
    Wenwu Wang$^{4}$,
    }
\secondlinename{	  
      Mark D. Plumbley$^{4}$, 
      Jun Yang$^{1,2}$
      }
\address{ $^{1}$Key Laboratory of Noise and Vibration Research, Institute of Acoustics, \\ 
Chinese Academy of Sciences, Beijing, China, \\
\{hujinbo, mingwu, feiran, yuziying, jyang\}@mail.ioa.ac.cn\\
$^{2}$University of Chinese Academy of Sciences, Beijing, China\\
$^{3}$Department of Intelligent Science, Xi'an Jiaotong Liverpool University, China, yin.k.cao@gmail.com\\
$^{4}$Centre for Vision, Speech and Signal Processing (CVSSP), University of Surrey, UK, \\
\{w.wang, m.plumbley\}@surrey.ac.uk
}

\begin{document}

\ninept
\maketitle

\begin{sloppy}

\begin{abstract}
For learning-based sound event localization and detection (SELD) methods, different acoustic environments in the training and test sets may result in large performance differences in the validation and evaluation stages. Different environments, such as different sizes of rooms, different reverberation times, and different background noise, may be reasons for a learning-based system to fail. On the other hand, acquiring annotated spatial sound event samples, which include onset and offset time stamps, class types of sound events, and direction-of-arrival (DOA) of sound sources is very expensive. In addition, deploying a SELD system in a new environment often poses challenges due to time-consuming training and fine-tuning processes. To address these issues, we propose Meta-SELD, which applies meta-learning methods to achieve fast adaptation to new environments. More specifically, based on Model Agnostic Meta-Learning (MAML), the proposed Meta-SELD aims to find good meta-initialized parameters to adapt to new environments with only a small number of samples and parameter updating iterations. We can then quickly adapt the meta-trained SELD model to unseen environments. Our experiments compare fine-tuning methods from pre-trained SELD models with our Meta-SELD on the Sony-TAU Realistic Spatial Soundscapes 2023 (STARSSS23) dataset. The evaluation results demonstrate the effectiveness of Meta-SELD when adapting to new environments.


\end{abstract}


\section{Introduction}
\label{sec:intro}

Sound event localization and detection (SELD) refers to detecting categories, presence, and spatial locations of different sound sources. SELD characterizes sound sources in a spatial-temporal manner. SELD was first introduced in Task 3 of the Detection and Classification of Acoustics Scenes and Events (DCASE) 2019 Challenge\cite{dcase2019task3}. After three iterations of Task 3 of the DCASE Challenge, types of data transform from computationally generated spatial recordings to real-scene recordings \cite{starss22}. 

SELD can be regarded as a Multi-Task Learning problem. Adavanne et al.\cite{Adavanne2018_JSTSP} proposed SELDnet for a joint task of sound event detection (SED) and regression-based direction-of-arrival (DOA) estimation. SELDnet is unable to detect homogeneous overlap, which refers to overlapping sound events of the same type but with different locations. The Event-Independent Network V2 (EINV2), with a track-wise output format and permutation invariant training, was proposed to tackle the homogeneous overlap detection problem\cite{cao2020event,cao2021,hu2022track}. Different from two outputs of SED and DOA in SELDnet and EINV2, the Activity-coupled Cartesian DOA (ACCDOA) approach merges two subtasks into a single task\cite{shimada2021accdoa, multiaccdoa}. The Cartesian DOA vectors contain the activity information of sound events in the ACCDOA method.

In practical SELD system deployment, unseen complex environments may lead to performance degradation. In the STARSS22 dataset\cite{starss22}, there are no duplicated recording environments in the training and validation sets. Our previous system submitted to Task 3 of the DCASE 2022 Challenge obtained the second rank in the team ranking\cite{hu22dw}. However, we found unsatisfactory generalization performance for fold4\_room2 recordings in the \textit{dev-test-tau} set of STARSS22\cite{hu22dw}. Experimental results show that class-dependent localization error $\mathrm{LE}_\mathrm{CD}$ is high and location-dependent F-score $\mathrm{F}_{\leq 20^\circ}$ is low, but class-dependent localization recall $\mathrm{LR}_\mathrm{CD}$ is high. This suggests there may be the weak localizing performance of our system in fold4\_room2. In addition, manually annotated spatial sound event recordings are very expensive. Taking the STARSS22 dataset for example\cite{starss22}, each scene was captured with a 32-channel spherical microphone array, a $360^\circ$ camera, a motion capture (mocap) system, and wireless microphones. Onset, offset, and class information of sound events were manually detected and classified by annotators through listening to wireless microphone recordings and watching video recordings, while positional annotations were extracted for each event by masking the tracker data with the temporal activity window of the event. In the end, $360^\circ$ video recordings are utilized to validate those annotations. This type of complex recording and annotation process means that large datasets of the annotated spatial recording might be expensive. 

Few-shot learning can act as a test bed for learning like humans, allowing a system to learn from small samples and reducing data gathering effort and computation\cite{wang2020fslsurvey}. Meta-learning, which facilitates few-shot learning, learns a general-purpose learning algorithm that generalizes across tasks and ideally enables each new task to be learned well from the task-distribution view\cite{meta-survey}. Meta-learning has advanced few-shot learning significantly in computer vision\cite{closerlook, mattersvison}. One of the most successful meta-learning algorithms is model-agnostic meta-learning (MAML)\cite{maml}. MAML tries to learn general initial parameters that can be rapidly adapted to another task. The method is model-agnostic and compatible with any model trained with gradient descent. It can be applicable to a variety of different learning problems, including classification, regression, and reinforcement learning. In audio signal processing, the meta-learning method has recently attracted interest as a way to solve few-shot learning problems recently. Meta-TTS\cite{meta-tts} is proposed to build personalized speech synthesis systems with few enrolled recordings of unseen users' voices using MAML. In \cite{meta-localization}, MAML is utilized to allow sound source localization models to adapt to different environments and conditions.

In this paper, we propose Meta-SELD, applying meta-learning to SELD models with activity-coupled Cartesian DOA (ACCDOA) representation\cite{shimada2021accdoa} to improve performance, especially in localization. We use MAML to find general initial parameters to minimize the loss across several tasks in Meta-SELD so that it can quickly adapt to an unseen environment. We take recordings in different environments as different tasks and aim to improve the performance of a specific unseen environment with a few samples recorded in the same environment. The experimental results demonstrate that Meta-SELD outperforms the fine-tuning method from the pre-trained SELD model in the STARSS23 dataset.

\section{Related Work}
\label{sec:related}

Activity-coupled Cartesian DOA (ACCDOA) representation\cite{shimada2021accdoa} assigns a sound event activity to the length of a corresponding Cartesian DOA. When inferring, the threshold is set for the length of class-wise Cartesian DOA vectors to determine whether an event class is active. In contrast to EINV2, the ACCDOA representation merges SED and DOA branches into a single branch, decreasing the model parameters and avoiding the necessity of balancing the loss measuring on the SED task and the DOA task.

The ACCDOA representation can not detect homogenous overlaps. Therefore, multi-ACCDOA which still contains a single branch and combines class-wise output format and track-wise output format, is proposed to overcome the problem\cite{multiaccdoa}. While each track in the track-wise output format of EINV2 only detects one event class and a corresponding location, each track in the multi-ACCDOA predicts activities and corresponding locations of all target classes. Auxiliary duplicating permutation invariant training (ADPIT) is also proposed to train each track of the multi-ACCDOA with original targets and duplicated targets, enabling each track to regard the same target as the single one. The multi-ACCDOA representation is shown in Fig. \ref{fig: multi-accdoa}. Its outputs are track-wise and class-wise Cartesian DOA vectors. Each vector length indicates the activity of the event. Besides the activity threshold, multi-ACCDOA employs angle thresholds to determine whether the predicted objects are the same or different.

\begin{figure}[tb]
  \centering
  \scalebox{1.}{\centerline{\includegraphics[width=\columnwidth]{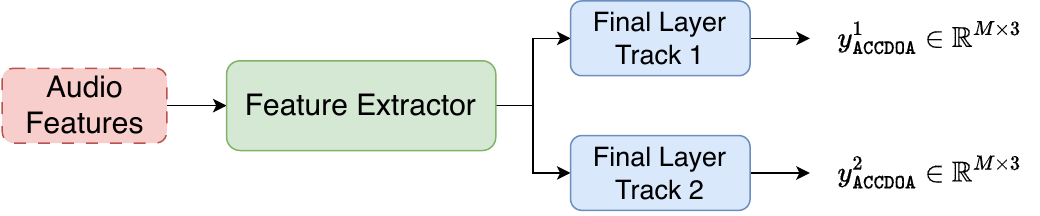}}}
  \caption{The multi-ACCDOA representation of the SELD model. There is no track dimension in the ACCDOA representation.}
  \label{fig: multi-accdoa}
\end{figure}

\section{Meta-SELD}

\subsection{The SELD model}
Without loss of generality, in this study, we adopt a simple Convolutional Recurrent Neural Network (CRNN) as our network, which is similar to the baseline of Task 3 of DCASE 2022 Challenge\cite{starss22} but with ACCDOA format. The network has three convolution blocks followed by a one-layer bidirectional gated recurrent unit (BiGRU). The network takes the concatenation of log-mel spectrograms and intensity vectors as input and predicts active sound events with corresponding Cartesian DOA vectors for each time step. The network architecture of CRNN is shown in Table \ref{table:CRNN}.

\begin{table}[h]
\centering
\caption{The network architecture of CRNN}
\label{table:CRNN}
\begin{adjustbox}{width=\columnwidth,center}
\begin{tabular}{c}
    \Xhline{3\arrayrulewidth}
    \multicolumn{1}{c}{ Log-mel spectrogram \& Intensity vectors} \\
    \hline
    \multicolumn{1}{c}{ $ ( \text{Conv2d }3 \times 3\ @ \ 32 \text{,  BatchNorm2d, ReLU}) \times 2 $, Avg Pooling $ 2 \times 2 $} \\
    \hline
    \multicolumn{1}{c}{ $ ( \text{Conv2d }3 \times 3\ @ \ 64 \text{,  BatchNorm2d, ReLU}) \times 2 $, Avg Pooling $ 2 \times 2 $} \\
    \hline
    \multicolumn{1}{c}{ $ ( \text{Conv2d }3 \times 3\ @ \ 128 \text{, BatchNorm2d, ReLU}) \times 2 $, Avg Pooling $ 2 \times 2 $} \\
    \hline
    \multicolumn{1}{c}{ $ ( \text{Conv2d }3 \times 3\ @ \ 256 \text{, BatchNorm2d, ReLU}) \times 2 $, Avg Pooling $ 1 \times 2 $} \\
    \hline
    \multicolumn{1}{c}{Global average pooling @ frequency} \\
    \hline
    \multicolumn{1}{c}{1-layer BiGRU of 128 hidden size, $256 \times 39$ linear layer, Tanh} \\
    \hline
    \multicolumn{1}{c}{Mean Square Error} \\
    \Xhline{3\arrayrulewidth}
    
\end{tabular}
\end{adjustbox}
\end{table}

\subsection{Meta-SELD training}

Given a model represented by a parameterized function $f_\Theta$ with parameters $\Theta$, MAML\cite{maml} learns the initial parameters $\Theta_{0}$ from general tasks $\mathcal{T}_i$ sampled from the training set $\mathcal{D}_\mathtt{train}$ and is expected to perform well on unseen tasks from the test set $\mathcal{D}_\mathtt{test}$ after a few iterations of parameters update with a small number of samples from the corresponding task. These initial parameters are very sensitive to being further optimized on a specific task. Each task $\mathcal{T}_i$ consists of a labeled support set $\mathcal{S}_i$ of $K$ samples and a labeled query set $\mathcal{Q}_i$ of $Q$ samples. A new task is expected to be quickly adapted with $K$ samples, which is known as $K$-shot learning. The loss function of MAML is defined as

\begin{equation}
\label{eq:MAML loss}
\mathcal{L} = \sum_{\mathcal{T}_i\sim p(\mathcal{T})} \mathcal{L}_{\mathcal{T}_i}(f_{\Theta})
\end{equation}
where $p(\mathcal{T})$, which is sampled from $\mathcal{D}_\mathtt{train}$, is a distribution over tasks that we want our model to be able to adapt to. In contrast to supervised deep learning methods, the objective of which is to find optimal parameters to minimize the loss function across all training samples, MAML tries to find generalized initial parameters for different tasks. MAML will then update the initial parameters after several iterations of training on data of new tasks.

There are two groups of parameters in the MAML algorithm, meta-parameters and adapt-parameters. In the meta-training phase, MAML starts with randomly initialized meta-parameters $\Theta$ and then adapts to a new specific task $\mathcal{T}_i$ with several update iterations using $\mathcal{S}_i$. The meta-parameters $\Theta$ become adapt-parameters $\Theta_i^\prime$:

\begin{equation}
\label{innerloop}
{\Theta}_i^\prime=\Theta-\alpha \nabla_{\Theta} \mathcal{L}_{\mathcal{T}_i}\left(f_{\Theta}, \mathcal{S}_i\right)
\end{equation}
where $\alpha$ is the adaptation learning rate for adapt-parameters updates. After updates across a batch of tasks, the meta-parameters are updated as:

\begin{equation}
\Theta=\Theta-\beta \nabla_{\Theta} \sum_{\mathcal{T}_i} \mathcal{L}_{\mathcal{T}_i}\left(f_{{\Theta}_i^\prime}, \mathcal{Q}_i\right)
\end{equation}
where $\beta$ is the meta step size. The loss $\mathcal{L}_{\mathcal{T}_i}$ is calculated by the parameterized function $f_{\Theta_i^\prime}$ on the query set $\mathcal{Q}_i$. After updating $\Theta$ on the query set, $\Theta$ will be used as initial parameters for the following meta-training steps.

We aim to adapt to an unseen environment with $K$ samples ($K$-shot). The objective of MAML is to find optimal initial parameters across several tasks, so we need to construct a set of tasks from the training set $\mathcal{D}_\mathtt{train}$. $\mathcal{D}_\mathtt{train}$ is split according to the different recording rooms. Audio clips recorded in different rooms belong to different tasks. We first sample a batch of tasks from all tasks and then sample $K+Q$ samples in each task, where $K$ samples for a support set $\mathcal{S}_i$ and $Q$ samples for a query set $\mathcal{Q}_i$. The overall training procedure of MAML is summarized in Algorithm \ref{alg:MAML}. Step 8 in Algorithm \ref{alg:MAML} is an inner-loop update for adapt-parameters, while Step 12 is outer-loop updates for meta-parameters.

\begin{algorithm}[t]
\caption{Meta-training of MAML for Meta-SELD}
\label{alg:MAML}
\begin{algorithmic}[1]
    \REQUIRE Distribution over all rooms $p(\mathcal{T})$, adaptation step size $\alpha$, meta step size $\beta$
    \STATE randomly initialize meta-parameters $\Theta$
    \WHILE{not done}
        \STATE Sample a batch of rooms $\mathcal{T}_i \sim p(\mathcal{T})$
        \FOR {each room $\mathcal{T}_i$}
            \STATE Sample disjoint examples $(\mathcal{S}_i, \mathcal{Q}_i)$ from $\mathcal{T}_i$
            \STATE Let $\Theta_{i,0} \gets \Theta$
            \FOR{gradient descent step $j:=0$ \textbf{to} $N-1$}
                \STATE Perform gradient descent to update adapt-parameters: $\Theta_{i,j+1} \gets \Theta_{i,j} - \alpha \nabla_{\Theta_i}\mathcal{L}_{\mathcal{T}_i}\left(\Theta_{i,j}, \mathcal{S}_i\right)$
            \ENDFOR
            \STATE Compute $\mathcal{L}_{\mathcal{T}_i}(f_{\Theta_{i,N}}, \mathcal{Q}_i)$
        \ENDFOR
        \STATE Perform gradient descent to update meta-parameters:\\ $\Theta \gets \Theta - \beta\nabla_{\Theta}\sum_{\mathcal{T}_i}\mathcal{L}_{\mathcal{T}_i}\left(f_{\Theta_{i,N}}, \mathcal{Q}_i\right)$
    \ENDWHILE
\end{algorithmic}
\end{algorithm}

\subsection{Meta-SELD test}
In the meta-testing phase, a specific unseen task $\mathcal{T}_j^\mathtt{test}$ created using $\mathcal{D}_\mathtt{test}$ is used. $\mathcal{T}_j^\mathtt{test}$ consists of a labeled support set $\mathcal{S}_j^\mathtt{test}$ of $K$ samples, and an unlabeled query set $\mathcal{Q}_j^\mathtt{test}$ of $Q$ samples. After training the model using well-trained parameter $\Theta$ from the meta-training phase as the initial parameters on $\mathcal{S}_j^\mathtt{test}$, we get updated parameters ${\Theta_j}^\prime$. We then use $f_{\Theta_j^\prime}$ to evaluate on $\mathcal{Q}_j^\mathtt{test}$.

The meta processes for testing and training are slightly different. Similar to the training, the test set $\mathcal{D}_\mathtt{test}$ is split according to the recording room of each audio clip. For clips of each room, we also chose $K$ samples for meta-test support set $\mathcal{S}_j^\mathtt{test}$ and all remaining samples for meta-test query set $\mathcal{Q}_j^\mathtt{test}$. After $N$ iterations of parameters update on $\mathcal{S}_j^\mathtt{test}$, the meta-parameters $\Theta$ are updated to ${\Theta_{j, N}}$. The final performance is evaluated on $\mathcal{Q}_j^\mathtt{test}$ with $f_{\Theta_{j, N}}$.

\section{Experiments}

\begin{table*}[ht]
    \centering
    \caption{The performance of the Meta-SELD and fine-tuning methods from pre-trained SELD models. Both two methods are evaluated in $\mathcal{Q}^\mathtt{test}_i$. Note that \textit{overall} scores of the fine-tuning method and Meta-SELD compute the fast adaptation performance of each individual room and then micro-average.}
    \label{tab:results}
    \vspace{-1mm}
    \resizebox{\textwidth}{!}{
        \begin{tabular}{l|ccc|ccc|ccc|ccc|ccc}
        \toprule
        \multicolumn{1}{c|}{\multirow{2}{*}{Room}} & &$\mathrm{ER}_{20^{\circ}} \downarrow$ & & & $\mathrm{F}_{20^{\circ}} \uparrow$ & & & $\mathrm{LE}_\mathrm{CD} \downarrow$ & & & $\mathrm{LR}_\mathrm{CD} \uparrow$ & & & $\mathcal{E}_{\mathtt{SELD}} \downarrow$ &\\
        & Pre-train & Fine-tune & Meta & Pre-train & Fine-tune & Meta & Pre-train & Fine-tune & Meta & Pre-train & Fine-tune & Meta & Pre-train & Fine-tune & Meta\\
        \midrule
        \text{fold3\_room4} & 0.624 & \textbf{0.574} & 0.603 & \textbf{44.5}\% & 40.4\% & 29.8\% & $17.8^\circ$ & $\boldsymbol{17.6^\circ}$ & $21.5^\circ$ & \textbf{64.6}\% & 61.2\% & 54.4\% & \textbf{0.408} & 0.414 & 0.470 \\
        \text{fold3\_room6} & 0.639 & 0.607 & \textbf{0.594} & 38.0\% & \textbf{40.5}\% & 40.4\% & $18.0^\circ$ & $\boldsymbol{17.2^\circ}$ & $17.4^\circ$ & \textbf{65.3}\% & 63.8\% & 61.1\% & 0.427 & \textbf{0.415} & 0.419 \\
        \text{fold3\_room7} & 0.610 & \textbf{0.606} & 0.660 & \textbf{31.1}\% & 30.7\% & 20.8\% & $23.6^\circ$ & $24.1^\circ$ & $\boldsymbol{22.5^\circ}$ & 59.9\% & \textbf{60.5}\% & 48.3\% & 0.458 & \textbf{0.457} & 0.523 \\
        \text{fold3\_room9} & 0.673 & \textbf{0.601} & 0.608 & 43.7\% & 46.6\% & \textbf{47.5}\% & $19.1^\circ$ & $18.6^\circ$ & $\boldsymbol{18.3^\circ}$ & \textbf{78.7}\% & 78.2\% & 73.3\% & 0.389 & \textbf{0.364} & 0.375 \\
        \text{fold3\_room12} & 0.685 & \textbf{0.659} & 0.689 & 28.0\% & 29.8\% & \textbf{33.0}\% & $26.8^\circ$ & $\boldsymbol{26.1^\circ}$ & $33.3^\circ$ & 43.1\% & 43.6\% & \textbf{46.3}\% & 0.531 & \textbf{0.518} & 0.520 \\
        \text{fold3\_room13} & 0.650 & 0.599 & \textbf{0.594} & 37.7\% & \textbf{39.4}\% & 36.1\% & $17.5^\circ$ & $16.9^\circ$ & $\boldsymbol{15.9^\circ}$ & \textbf{50.9}\% & 48.8\% & 37.1\% & 0.465 & \textbf{0.453} & 0.488 \\
        \text{fold3\_room14} & 0.633 & \textbf{0.582} & 0.613 & \textbf{40.2}\% & 37.4\% & 28.6\% & $\boldsymbol{23.2^\circ}$ & $23.7^\circ$ & $24.8^\circ$ & \textbf{55.3}\% & 54.0\% & 47.2\% & 0.452 & \textbf{0.450} & 0.498 \\
        \text{fold3\_room21} & 0.757 & 0.750 & \textbf{0.735} & 19.3\% & \textbf{21.6}\% & 18.9\% & $20.5^\circ$ & $\boldsymbol{18.9^\circ}$ & $20.6^\circ$ & 39.3\% & 31.4\% & \textbf{43.8}\% & 0.571 & 0.581 & \textbf{0.556} \\
        \text{fold3\_room22} & 0.850 & 0.818 & \textbf{0.800} & 11.4\% & 12.8\% & \textbf{16.7}\% & $31.6^\circ$ & $29.5^\circ$ & $\boldsymbol{29.0^\circ}$ & 45.6\% & 43.8\% & \textbf{48.8}\% & 0.614 & 0.604 & \textbf{0.577} \\
        \text{fold4\_room2} & 0.809 & 0.774 & \textbf{0.753} & 6.2\% & 8.2\% & \textbf{15.4}\% & $47.8^\circ$ & $41.3^\circ$ & $\boldsymbol{33.0^\circ}$ & 72.4\% & 72.4\% & \textbf{75.7}\% & 0.572 & 0.550 & \textbf{0.506} \\
        \text{fold4\_room8} & 0.716 & 0.716 & \textbf{0.702} & 31.7\% & \textbf{33.6}\% & 30.7\% & $22.5^\circ$ & $\boldsymbol{21.0^\circ}$ & $23.2^\circ$ & \textbf{54.0}\% & 49.4\% & 49.4\% & \textbf{0.496} & 0.501 & 0.507 \\
        \text{fold4\_room10} & 0.792 & 0.708 & \textbf{0.651} & 36.3\% & \textbf{41.7}\% & 35.8\% & $23.8^\circ$ & $21.5^\circ$ & $\boldsymbol{20.2^\circ}$ & 66.1\% & 72.0\% & \textbf{78.2}\% & 0.475 & 0.423 & \textbf{0.406} \\
        \text{fold4\_room15} & 0.582 & 0.563 & \textbf{0.539} & 33.3\% & 33.5\% & \textbf{43.4}\% & $16.5^\circ$ & $\boldsymbol{15.5^\circ}$ & $19.3^\circ$ & 42.8\% & 42.6\% & \textbf{59.0}\% & 0.478 & 0.472 & \textbf{0.406} \\
        \text{fold4\_room16} & 0.601 & \textbf{0.584} & 0.607 & 39.8\% & \textbf{40.5}\% & 34.3\% & $21.7^\circ$ & $21.9^\circ$ & $\boldsymbol{21.6^\circ}$ & \textbf{55.1}\% & 54.9\% & 48.7\% & 0.443 & \textbf{0.438} & 0.474 \\
        \text{fold4\_room23} & 0.813 & 0.746 & \textbf{0.676} & 25.4\% & 26.5\% & \textbf{31.8}\% & $26.2^\circ$ & $\boldsymbol{24.9^\circ}$ & $25.8^\circ$ & 40.4\% & 43.6\% & \textbf{47.3}\% & 0.575 & 0.546 & \textbf{0.507} \\
        \text{fold4\_room24} & 0.828 & \textbf{0.779} & 0.782 & 26.2\% & 25.7\% & \textbf{30.8}\% & $\boldsymbol{19.4^\circ}$ & $19.7^\circ$ & $24.4^\circ$ & 41.0\% & \textbf{43.6}\% & 42.7\% & 0.566 & 0.549 & \textbf{0.546} \\
        \midrule
        \multicolumn{1}{c|}{\text{Overall}} & 0.707 & 0.677 & \textbf{0.672} & 23.0\% & 24.2\% & \textbf{26.0}\% & $22.8^\circ$ & $22.3^\circ$ & $\boldsymbol{21.9^\circ}$ & 39.5\% & 40.2\% & \textbf{41.0}\% & 0.552 & 0.539 & \textbf{0.531} \\
        \bottomrule
        \end{tabular}
        }
\vspace{-4mm}
\end{table*}

\subsection{Dataset}
There are 16 different recording rooms in total in the development set of the STARSS23 dataset, including nine recording rooms in \textit{dev-train-set} and seven recordings rooms in \textit{dev-test-set}. The development set of STARSS23, which contains roughly 7.5 hours of recordings, has less data than the development set in DCASE 2021, which contains roughly 13 hours of synthetic recordings\cite{dcase2021task3}. Considering the complexity of the real-scene environment, we use additional datasets to improve the performance. We generated simulated data using the generator code provided by DCASE\footnote{https://github.com/danielkrause/DCASE2022-data-generator}. We synthesize multi-channel spatial recordings by convolving monophonic sound event examples with multi-channel Spatial Room Impulse Responses (SRIRs). Samples of sound events are selected from AudioSet\cite{gemmeke2017audio} and FSD50K\cite{fonseca2021fsd50k}, based on the affinity of the labels in those datasets to target classes in STARSS23. PANNs\cite{kong2020panns} are then employed to clean the selection of the clips. We use pre-trained PANNs to infer these clips and select high-quality clips based on output probability above 0.8. We extracted SRIRs from the TAU Spatial Room Impulse Response Database (TAU-SRIR DB)\footnote{https://zenodo.org/record/6408611}, which contains SRIRs captured in 9 rooms at Tampere University. It was used for official synthetic datasets in DCASE 2019-2021\cite{dcase2019task3,dcase2020task3,dcase2021task3}.

The 2700 1-minute audio clips that we synthesized using the abovementioned SRIRs from 9 rooms are used for $\mathcal{D}_\mathtt{train}$, and all of \textit{dev-set} of STARSS23, recorded in 16 rooms, are used for $\mathcal{D}_\mathtt{test}$.

\subsection{Experimental setup}
 The sampling rate of the dataset is 24 kHz. We extracted 64-dimensional log mel spectrograms from four-channel first-order ambisonics (FOA) signals with a Hanning window of 1024 points, and a hop size of 320. Each audio clip is segmented to a fixed length of five seconds with no overlap for training and inference.

 In the meta-training phase, the training set and test set are divided into 9 tasks and 16 tasks, respectively, corresponding to 9 rooms and 16 rooms. We first sample a batch of rooms randomly and then sample a batch of examples from each of the rooms. The batch of samples of each room constructs a task, and a part of the samples are support samples while the remaining samples are query samples. The batch size of rooms and samples is 4 and 64, respectively. A batch of samples contains 30 support samples and 34 query samples. In the meta-test phase, we sort the audio clips according to the filename, and select the first 30 samples of recordings of each room as samples from the support set $\mathcal{S}_j^\mathtt{test}$. The remaining samples of each room are as samples from the test set $\mathcal{Q}_j^\mathtt{test}$. The AdamW optimizer is used for updates of meta-parameters of MAML, while the SGD optimizer is used to update adapt-parameters. The meta step size $\beta$ begins with 0.001 in the first 100 epochs out of 150 epochs in total and is then decreased by 10\% every 20 epochs. The adaptation step size and the number of update iterations are always kept at 0.01 and 5, respectively. 

To demonstrate the effectiveness of Meta-SELD, we compare Meta-SELD with the fine-tuning method from the pre-trained SELD model. Firstly, we train a SELD model with AdamW optimizer in $\mathcal{D}_\mathtt{train}$ from scratch. The learning rate is 0.0003 for the first 70 epochs and then decreases to 0.00003 for the following 20 epochs. Secondly, we initialize the parameters from the previously trained SELD model and then use $\mathcal{S}^\mathtt{test}_i$ and $\mathcal{Q}^\mathtt{test}_i$ as the training set and the test set of the $i$-th room to fine-tune. Similar to the process of the adapt-parameters updates in MAML, the SGD optimizer with a step size of 0.01 and update iterations of 5 are used for fine-tuning.

A joint metric of localization and detection\cite{mesaros2019joint,overviewofDCASE} is used: location-dependent F-score ($\mathrm{F}_{\leq 20^\circ}$) and error rate ($\mathrm{ER}_{\leq 20^\circ}$), and class-dependent localization recall ($\mathrm{LR}_\mathrm{CD}$) and localization error ($\mathrm{LE}_\mathrm{CD}$). $\mathrm{F}_{\leq 20^\circ}$ and $\mathrm{ER}_{\leq 20^\circ}$ consider true positives predicted under a spatial threshold $20^\circ$  from the ground truth. $\mathrm{LE}_\mathrm{CD}$ and $\mathrm{LR}_\mathrm{CD}$ are computed for localization predictions in the case that the types of sound events are predicted correctly. A macro-average of $\mathrm{F}_{\leq 20^\circ}$, $\mathrm{LR}_\mathrm{CD}$ and $\mathrm{LE}_\mathrm{CD}$ is used.

We use an aggregated SELD metric which was computed as
\begin{equation}
{\mathcal{E}_\mathtt{SELD}}=\frac{1}{4}\left[\mathrm{ER}_{\leq 20^{\circ}}+\left(1-\mathrm{F}_{\leq 20^{\circ}}\right)+\frac{\mathrm{LE}_{\mathrm{CD}}}{180^{\circ}}+\left(1-\mathrm{LR}_{\mathrm{CD}}\right)\right].
\end{equation}

\subsection{Experimental results}
Table \ref{tab:results} shows the performance of the Meta-SELD method compared with the fine-tuning method from the pre-trained SELD models. The pre-trained SELD models are trained without using samples from $\mathcal{D}_\mathtt{test}$.

According to the last row of Table \ref{tab:results}, the \textit{overall} score, which is a micro average across all rooms, shows that all of $\mathrm{ER}_{\leq 20^{\circ}}$, $\mathrm{F}_{\leq 20^{\circ}}$, $\mathrm{LE}_{\mathrm{CD}}$, and $\mathrm{LR}_{\mathrm{CD}}$ are improved using Meta-SELD compared with the fine-tuning method. We observe a drop in $\mathcal{E}_\mathtt{SELD}$ in fold3\_room4 and fold4\_room8 even though some new samples of unseen environments are used for training. This may be due to the fact that the new samples do not have valid information for training. We also observe the Meta-SELD method improves $\mathcal{E}_\mathtt{SELD}$ by a large margin in fold3\_room22, fold4\_room2, and fold4\_room23 where the pre-trained model has poor performance across all rooms. 
Specifically, $\mathrm{ER}_{\leq 20^{\circ}}$, $\mathrm{F}_{\leq 20^{\circ}}$, and $\mathrm{LR}_{\mathrm{CD}}$ of fold3\_room22 and fold4\_room23 outperform other methods. Meta-SELD mainly improves the performance of SED in fold3\_room22 and fold4\_room23. All metrics of fold4\_room2 are improved in Meta-SELD compared with the fine-tuning method, especially in DOA estimation. In fold4\_room2, all of the pre-trained model, the fine-tuning method, and Meta-SELD achieve $\mathrm{LR}_{\mathrm{CD}}$ of over 70\%, but $\mathrm{LE}_{\mathrm{CD}}$ of three methods is always high compared with $\mathrm{LE}_{\mathrm{CD}}$ of other rooms. Meta-SELD decreases $14.8^\circ$ and $8.3^\circ$ of $\mathrm{LE}_{\mathrm{CD}}$ compared with the pre-trained model and the fine-tuning method in fold4\_room2, hence directly leading to the increase of $\mathrm{F}_{\leq 20^{\circ}}$ and the decrease of $\mathrm{ER}_{\leq 20^{\circ}}$. However, performance degradation happens in fold3\_room4, fold3\_room7, fold3\_room14, and fold4\_room16, where Meta-SELD has the worst metric scores. There is no significant change in $\mathrm{LE}_{\mathrm{CD}}$, and the decline in SED performance is the main factor. One of the possible reasons for this observation could be that there are some conflicts in optimizing Meta-SELD across a batch of rooms.

Experimental results demonstrate that Meta-SELD can find better initial parameters across a batch of tasks than the fine-tuning method, especially in rooms where the pre-trained model and the fine-tuning method perform worse. Meta-SELD reduces the risk of overfitting when using a small number of samples, which usually happens in the fine-tuning method. 

\section{Conclusion}
In this paper, we presented Meta-SELD, which employed Model-Agnostic Meta-Learning (MAML) to the sound event localization and detection task to achieve fast adaptation to unseen environments. The method only utilizes a small number of samples and a few update iterations of training. We use the STARSS23 dataset and synthesized 2700 1-minute samples that are convolved using monophonic sound event clips with multi-channel spatial room impulse responses. The sound event clips are extracted from FSD50K and AudioSet and are further filtered by the PANNs model through a probability threshold. The SRIRs used are from TAU-SRIR DB. Our methods are trained on synthetic datasets and evaluated on all development sets of the STARSS23 dataset. Audio clips recorded from the same room or synthesized using SRIRs collected from the same room are regarded as the same task for MAML. The experimental results show that the Meta-SELD method improves $\mathcal{E}_\mathtt{SELD}$ significantly in those rooms where both the pre-trained model and the fine-tuning method perform unsatisfactorily. The overall score demonstrates that the Meta-SELD method outperforms the fine-tuning method on average. 

\section{Acknowledgement}
\label{sec:ack}
This work was supported in part by Grant ``XJTLU RDF-22-01-084'', UK Engineering and Physical Sciences Research Council (EPSRC) Grant EP/T019751/1 ``AI for Sound (AI4S)''. For the purpose of open access, the authors have applied a Creative Commons Attribution (CC BY) licence to any Author Accepted Manuscript version arising.


\bibliographystyle{IEEEtran}
\bibliography{refs}

\end{sloppy}
\end{document}